\documentclass[aps,prd,preprint,groupedaddress,showpacs]{revtex4-1}

\usepackage{amsmath,amssymb}
\usepackage[dvips]{graphicx}
\usepackage{braket}

\begin{document}

\preprint{}

\title{The Aharonov-Bohm Effect on Entanglement Entropy in Conformal Field Theory}

%Entanglement Entropy and the Aharonov-Bohm effect in Conformal Field Theory


\author{Noburo Shiba}
%\email{shiba@het.phys.sci.osaka-u.ac.jp} 
\email{nshiba@fas.harvard.edu}
%{Your e-mail address}
%\thanks{}
\affiliation{Center for the Fundamental Laws of Nature,\\
Harvard University, Cambridge, MA 02138 USA}

%\affiliation{Yukawa Institute for Theoretical Physics (YITP),\\
%Kyoto University, Kyoto 606-8502, Japan}


\date{\today}

\begin{abstract}
We consider the Aharonov-Bohm effect on entanglement entropy for one interval in (1+1) dimensional conformal field theory on a one dimensional ring. 
The magnetic field is confined inside the ring, i.e. there is a Wilson loop %which is proportional to the Aharonov-Bohm phase 
on the ring. 
%in (1+1) dimensional conformal field theory. 
%The subsystem is one interval on the ring. 
The Aharonov-Bohm phase factor which is proportional to the Wilson loop is represented as insertion of twist operators. 
We compute exactly the R\'{e}nyi entropy from a four point function of twist operators in a free charged scalar field. 
\end{abstract}

%\keywords{}

\maketitle

\section{Introduction}
The entanglement entropy in the quantum field theory plays 
 important roles in many fields of physics including the string theory \cite{RT, Fa, Sw, NRT, MT, RaT,  Sh3, Sh4, Sh7, Sh8, Sh9, Sh10}, 
condensed matter physics \cite{LW, KP, CC}, lattice gauge theories \cite{GST, Sh6}, and the physics of the black hole \cite{Bombelli:1986rw, Sr, SU, Ka, Sh1, Sh2}. 
The entanglement entropy is a useful quantity which characterize 
quantum properties of  given states. 
%For example, the entanglement entropy of ground states 
%follows the area law \cite{Bombelli:1986rw, Sr, Ereview, La} if we consider a local quantum field theory with a UV fixed point,
%while non-local field theories \cite{ShTa, Ka} or QFTs with fermi surfaces \cite{FS1} at UV cut off scale can violate the
%area law. 

For a given density matrix $\rho$ of the total system, 
the entanglement entropy of the subsystem $\Omega$ is defined as 
\begin{equation}
S_{\Omega} =-\mathrm{Tr} \rho_{\Omega} \ln \rho_{\Omega}, 
\end{equation}
where $\rho_{\Omega} =\mathrm{Tr}_{\Omega^{c}}\rho$ 
is the reduced density matrix of the subsystem $\Omega$ 
and $\Omega^c$ is the complement of $\Omega$. 
%The useful generalization of the entanglement entro
%$S_{\Omega}$ can be computed as limit as $n\rith$
The R\'{e}nyi entropy $S_{\Omega}^{(n)}$ is defined as 
\begin{equation}
S_{\Omega}^{(n)} = \dfrac{1}{1-n} \ln \mathrm{Tr} \rho_{\Omega}^{n} .
\end{equation}
The limit $n\rightarrow 1$ coincides with the entanglement entropy 
$\lim_{n=1} S_{\Omega}^{(n)}=S_{\Omega}$.

On the other hand, 
the Aharonov-Bohm (AB) effect is a fundamental quantum phenomenon in which an electrically charged particle is affected by an electromagnetic potential $A_{\mu}$, despite being confined to a region in which both the magnetic and electric field are zero.

\begin{figure}
\includegraphics[width=6.5cm,clip]{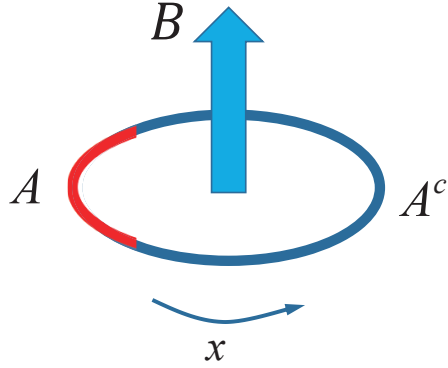}%
\caption{One dimensional ring studied in this paper. The circumference of the ring is $L$ and the subsystem $A$ is one interval whose length is $l$. The space coordinate $x$ has the periodicity $x \sim x+L$.  The magnetic field is confined inside the ring and induces the Aharonov-Bohm phase on the ring. \label{ABphase}}
\end{figure}

In this paper, we consider the dependence of entanglement entropy with the AB phase. 
In particular, we consider (1+1) dimensional conformal field theory on a one dimensional ring and study how the entanglement entropy for one interval on the ring is affected by a magnetic field enclosed by it (see Fig.\ref{ABphase}). 
%The Aharonov-Bohm phase factor which is proportional to the Wilson loop is represented as insertion of twist operators. 
The Aharonov-Bohm phase factor can be represented as a twisted boundary condition by a gauge transformation. 
Thus, the twisted boundary condition is represented as insertion of twist operators.  
We compute exactly the R\'{e}nyi entropy from a four point function of twist operators in a free charged scalar field. 

The Aharonov-Bohm effect on entanglement entropy was studied in \cite{ABC}. 
In \cite{ABC}, entanglement entropy for free charged scalar and Dirac fields in an annular strip on two dimensional cylinder was studied.  
%The Aharonov-Bohm phase factor is represented as a twisted boundary condition by a gauge transformation. 
Entanglement entropy in quantum field theories with twisted boundary conditions was studied in \cite{CCDSLMV, MFS, WWS, CWFF}.

\section{The Aharonov-Bohm Effect on Entanglement Entropy in 2d CFT}
\begin{figure}
\includegraphics[width=10cm,clip]{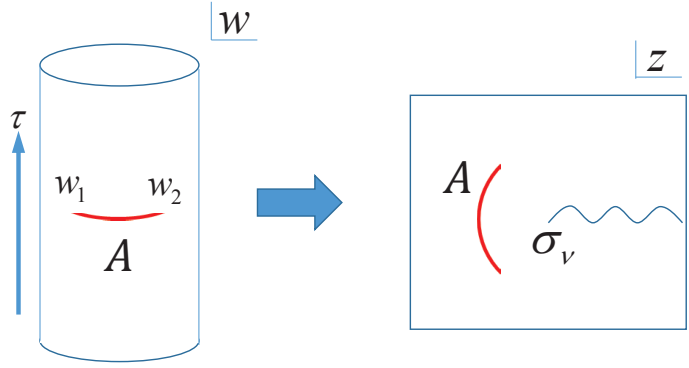}%
\caption{The Euclidean path integral for $\textrm{Tr}[\rho_A^n]$ in $w$ and $z$ coordinates. In $z$-plane, the boundary condition (\ref{bc2}) can be expressed by inserting twist operators $\sigma_\nu$ and $\sigma_{1-\nu}$ at $z=0$ and $z=\infty$.  \label{ABpath}}
\end{figure}

%We consider the Aharonov-Bohm Effect on entanglement entropy for one interval on the ring
We consider (1+1) dimensional conformal field theory on a one dimensional ring whose circumference is $L$. 
The space coordinate $x$ has the periodicity $x \sim x+L$. 
We analyze a complex scalar field, $\phi$, charged with respect to an external gauge field, $A_\mu$, which is pure gauge on the ring.   
We assume that $\phi(x)$ has the periodicity $\phi(x)=\phi(x+L)$. 
We choose a constant gauge field, $A_x$, in the $x$ direction. 
We can eliminate it by a gauge transformation 
\begin{equation}
\phi'(x)=e^{-iq \int^x dx' A_x} \phi (x) ,  \label{gauge}
\end{equation}
where $q$ is a charge of $\phi$. 
The scalar field has now the following boundary condition
\begin{equation}
\phi'(x+L)=e^{-iq A_x L} \phi' (x) \equiv e^{-i 2 \pi \nu} \phi' (x) , \label{bc1}
\end{equation}
where we defined $\nu \equiv \frac{q}{2\pi} \Phi =\frac{q}{2\pi} L A_x$ and $\Phi \equiv \oint dx' A_x =L A_x$. 
The integral $\Phi$ is the magnetic flux inside the ring  and $\nu$ is the Aharonov-Bohm (AB) phase. 
Now we consider the R\'{e}nyi entropy, $S_{A}^{(n)} = \frac{1}{1-n} \ln \mathrm{Tr} \rho_{A}^{n}$, for one interval whose length is $l$. 
We compute the R\'{e}nyi entropy by using the replica method and the Euclidean path integral \cite{CC}. 
The Euclidean coordinate is $w=x+i \tau$, where $x$ is the space coordinate and has periodicity $x\sim x+L$, and $\tau$ is the Euclidean time ($-\infty<\tau<\infty$). 
We define the subsystem $A$ to be the interval $x_1 \leq x \leq x_2$, $\tau_1=\tau_2=0$,  $x_2-x_1=l$, where $w_1=x_1$ and $w_2=x_2$ are endpoints of the interval. 
The R\'{e}nyi entropy is expressed as the expectation value of twist operators, 
\begin{equation}
\textrm{Tr}[\rho_A^n] =
\langle \mathcal{T}_n (w_1, \bar{w}_1)  \tilde{\mathcal{T}}_n (w_2, \bar{w}_2) \rangle_{\nu} , \label{vev w}
\end{equation}
where $\langle \dots \rangle_{\nu}$ is the expectation value under the boundary condition (\ref{bc1}), and $\mathcal{T}_n$ and $\tilde{\mathcal{T}}_n$ are the twist operators whose action is  
\begin{equation}
\mathcal{T}_n : \phi_{i} ' \rightarrow \phi_{i+1}' ~~ (\mathrm{mod}~ n), ~~~
\tilde{\mathcal{T}}_n  : \phi_{i+1}' \rightarrow \phi_{i}' ~~ (\mathrm{mod}~ n),
\end{equation}
here $\phi_i$ denotes the $i$th replica field. 
To compute (\ref{vev w}), we use the conformal map $z=e^{-i\frac{2\pi}{L} w}$. 
From (\ref{bc1}), the scalar field in the $z$ plane has the following boundary condition,  
\begin{equation}
\phi'(e^{i2\pi}z, e^{-i2\pi} \bar{z})= e^{i 2 \pi \nu} \phi' (z,\bar{z}) . \label{bc2}
\end{equation}
The boundary condition (\ref{bc2}) can be expressed by inserting twist operators $\sigma_\nu$ and $\sigma_{1-\nu}$ at $z=0$ and $z=\infty$ (See Fig.\ref{ABpath}).
The action of $\sigma_{\alpha}$ is
\begin{equation}
\sigma_\alpha : \phi_{i} ' \rightarrow e^{i2\pi \alpha} \phi_{i}' .
\end{equation}
Thus, we rewrite  $\textrm{Tr}[\rho_A^n]$ in (\ref{vev w}) as 
\begin{equation}
\textrm{Tr}[\rho_A^n] = 
\left| \frac{dw_1}{dz_1} \right|^{-2h_n}  \left| \frac{dw_2}{dz_2} \right|^{-2h_n} 
\frac{\langle \mathcal{T}_n (z_1)  \tilde{\mathcal{T}}_n (z_2) 
 \mathcal{\sigma}_\nu (0)  \mathcal{\sigma}_{1-\nu} (\infty) \rangle}{\langle \mathcal{\sigma}_\nu (0)  \mathcal{\sigma}_{1-\nu} (\infty) \rangle} ,\label{vev z}
\end{equation}
where $z_{1,2}=e^{-i\frac{2\pi}{L} w_{1,2}}$ and $h_n=\frac{c}{24} (n-1/n)$ is the conformal weight of  $\mathcal{T}_n$ and $ \tilde{\mathcal{T}}_n$, 
here $c$ is the central charge.

\section{Charged free scalar field}
We apply (\ref{vev z}) to a free charged scalar field. 
For the free scalar field, it is useful to use the following Fourier transformation, 
\begin{equation}
\tilde{\phi_k}\equiv \sum_{j=0}^{n-1} e^{i2\pi \frac{k}{n}j} \phi_{j}' .
\end{equation}
For free fields, the Fourier transformation diagonalizes the action of $\mathcal{T}_n$, $ \tilde{\mathcal{T}}_n$ and $\mathcal{\sigma}_{\alpha}$ simultaneously, 
\begin{equation}
\mathcal{T}_n : \tilde{\phi_{k}} \rightarrow e^{i2\pi \frac{k}{n}} \tilde{\phi_{k}}, ~~
\tilde{\mathcal{T}}_n : \tilde{\phi_{k}} \rightarrow e^{-i2\pi \frac{k}{n}} \tilde{\phi_{k}}, ~~
\sigma_\alpha : \tilde{\phi_{k}} \rightarrow e^{i2\pi \alpha} \tilde{\phi_{k}} .
\end{equation}
Thus, the four point function of the twist operators in (\ref{vev z}) become 
\begin{equation}
\begin{split}
\frac{\langle \mathcal{T}_n (z_1)  \tilde{\mathcal{T}}_n (z_2) 
 \mathcal{\sigma}_\nu (0)  \mathcal{\sigma}_{1-\nu} (\infty) \rangle}{\langle \mathcal{\sigma}_\nu (0)  \mathcal{\sigma}_{1-\nu} (\infty) \rangle}
&=\prod_{k=1}^{n-1}
\frac{\langle \mathcal{\sigma}_{k/n} (z_1)  \mathcal{\sigma}_{1-k/n} (z_2) 
 \mathcal{\sigma}_\nu (0)  \mathcal{\sigma}_{1-\nu} (\infty) \rangle}{\langle \mathcal{\sigma}_\nu (0)  \mathcal{\sigma}_{1-\nu} (\infty) \rangle} \\
&=\prod_{k=1}^{n-1}
\frac{\langle \mathcal{\sigma}_{k/n} (0)  \mathcal{\sigma}_{1-k/n} (x) 
 \mathcal{\sigma}_\nu (1)  \mathcal{\sigma}_{1-\nu} (\infty) \rangle}{\langle  \mathcal{\sigma}_\nu (1)  \mathcal{\sigma}_{1-\nu} (\infty) \rangle} \\
\end{split} \label{vev x}
\end{equation}
where we used the conformal map $f(z)=1-z/z_1$ and $x=f(z_2)=1-e^{-i\frac{2\pi}{L}l}$ is the cross ratio of the four points ( $|x|=2|\sin \frac{\pi l}{L}|$ ). 
From (\ref{vev z}) and (\ref{vev x}), we obtain the R\'{e}nyi entropy, 
\begin{equation}
S_A^{(n)}=\frac{1}{1-n} \ln \textrm{Tr}[\rho_A^n] = \frac{1}{1-n} \sum_{k=0}^{n-1} \ln \frac{\langle \mathcal{\sigma}_{k/n} (0)  \mathcal{\sigma}_{1-k/n} (x) 
 \mathcal{\sigma}_\nu (1)  \mathcal{\sigma}_{1-\nu} (\infty) \rangle}{\langle  \mathcal{\sigma}_\nu (1)  \mathcal{\sigma}_{1-\nu} (\infty) \rangle} \label{ree1}
\end{equation}
where we used $\left| \frac{dw_{1,2}}{dz_{1,2}} \right| =\frac{2\pi}{L} $ and omitted the irrelevant constant. 

%The four point function of the twist fields was calculated by Knizhnik \cite{Kn}.  
The four point function of twist operators also appear in the calculation of R\'{e}nyi entropy of two disjoint intervals in free scalar field theory \cite{CCT}. 
In the case of two disjoint intervals, the necessary four point function is $\langle \mathcal{\sigma}_{k/n} (0)  \mathcal{\sigma}_{1-k/n} (x) 
 \mathcal{\sigma}_{k/n} (1)  \mathcal{\sigma}_{1-k/n} (\infty) \rangle$. 
In our case, we need the more general four point function $\langle \mathcal{\sigma}_{k_1/n} (0)  \mathcal{\sigma}_{1-k_1/n} (x) 
 \mathcal{\sigma}_{k_3/n} (1)  \mathcal{\sigma}_{1-k_3/n} (\infty) \rangle $. 
In the following we will use the results of the four point function of the twist operators by Knizhnik  \cite{Kn}. 
We give derivation and different expression of $\langle \mathcal{\sigma}_{k_1/n} (0)  \mathcal{\sigma}_{1-k_1/n} (x) 
 \mathcal{\sigma}_{k_3/n} (1)  \mathcal{\sigma}_{1-k_3/n} (\infty) \rangle $ by another method  in the Appendix B. 
Note that there are a series of papers from late eighties about conformal field theories on orbifold (e.g. \cite{DFMS, BR, ADGN}) that are probably useful for more complicated cases. 

The four point function of the twist operators is given by (see equations (7.22) and (7.28) in \cite{Kn}), 
\begin{equation}
\begin{split}
\langle \mathcal{\sigma}_{k_1/n} (0)  \mathcal{\sigma}_{k_2/n} (x) 
 \mathcal{\sigma}_{k_3/n} (1)  \mathcal{\sigma}_{2-(k_1+k_2+k_3)/n} (\infty) \rangle 
&= \kappa^2 (Z Z_{*})^{-1/2}, \\  \label{four point 1}
\end{split}
\end{equation}
\begin{equation}
\begin{split}
Z_{*}(\{ k_i \}|x) & = |x|^{2 k_1 k_2 / n^2}  |1-x|^{2 k_2 k_3 / n^2} 
I (-k_1 /n, -k_2/n, -k_3/n, x),  \\  \label{Z1}
\end{split}
\end{equation}
\begin{equation}
\begin{split}
Z(\{ k_i \}|x) &= Z_{*}(\{ n- k_i \}|x),  \\  \label{Z2}
%& = |x|^{2 k_1 k_2 / n^2}  |1-x|^{2 k_2 k_3 / n^2} 
%I (-k_1 /n, -k_2/n, -k_3/n, x)  \\
\end{split}
\end{equation}
\begin{equation}
\begin{split}
& I (a, b, c, x) = \int d^2 z |z|^{2a}  |z-x|^{2b}  |z-1|^{2c},  \\  \label{I1}
\end{split}
\end{equation}
where $\kappa$ is a constant and $\int d^2 z= \int_{-\infty}^{\infty} d \mathrm{Re} z ~ \int_{-\infty}^{\infty} d \mathrm{Im} z$.  
%Furthermore, we give derivation of $\langle \mathcal{\sigma}_{k_1/n} (0)  \mathcal{\sigma}_{1-k_1/n} (x) 
% \mathcal{\sigma}_{k_3/n} (1)  \mathcal{\sigma}_{1-k_3/n} (\infty) \rangle $ and the different expression by another method in the Appendix B. 
Note that $\mathcal{\sigma}_{k/n} $ in \cite{Kn} (and in (\ref{four point 1}))  is normalized as $\langle \mathcal{\sigma}_{k/n}(0) \mathcal{\sigma}_{1-k/n}(x)\rangle =|x|^{-2 \frac{k}{n} \left(1-\frac{k}{n} \right)}$. 
On the other hand, $\mathcal{\sigma}_{k/n} $ in (\ref{ree1}) is normalized as $\langle \mathcal{\sigma}_{k/n}(0) \mathcal{\sigma}_{1-k/n}(x)\rangle =\left(|x|/\epsilon \right)^{-2 \frac{k}{n} \left(1-\frac{k}{n} \right)}$, here $\epsilon \equiv a/L$ and $a$ is the UV cutoff length. 
The latter normalization is usually used in calculation of R\'{e}nyi entropy and gives the correct UV cutoff dependence of R\'{e}nyi entropy. 
The integral $I(a, b, c, x)$ is calculated in the Appendix A. 
Note that %different expression is written in the Appendix in \cite{Kn}. 
the expression of $I(a, b, c, x)$ in the Appendix in \cite{Kn} is not useful when $a+b=-1$ and we give a different expression which is useful when $a+b=-1$ in (\ref{I2}).
Thus, from (\ref{ree1})-(\ref{I1}), we obtain the R\'{e}nyi entropy, %for the free charged scalar field, 
\begin{equation}
\begin{split}
&S_A^{(n)} = \frac{1}{1-n} \sum_{k=1}^{n-1} \ln \Bigl[ \left( \frac{1}{\epsilon}\right)^{-2 \frac{k}{n} \left(1-\frac{k}{n} \right)} \kappa^2 (Z Z_{*}(k_1=k, k_2=n-k, k_3=n\nu))^{-1/2} \Bigr]  \\
& = \frac{1}{1-n} \sum_{k=1}^{n-1} \ln \Bigl[ \left( \frac{1}{\epsilon}\right)^{-2 \frac{k}{n} \left(1-\frac{k}{n} \right)} \kappa^2 (|x|^{4\frac{k}{n} \left(1-\frac{k}{n} \right)} |1-x|^{4\frac{k}{n} \left(1-\nu \right)} I\left(\tfrac{k}{n}-1, -\tfrac{k}{n}, \nu-1, x\right)^2)^{-1/2}  \Bigr]  \\
&= \frac{1}{1-n} \sum_{k=1}^{n-1} \ln  \Bigl[ \kappa^2
\left( \frac{|x|}{ \epsilon} \right)^{-2 \frac{k}{n} \left(1-\frac{k}{n} \right)} 
  |1-x|^{-2 \frac{k}{n} \left(1-\nu \right)} \\ 
& \times \Bigl(   \frac{\Gamma (1-\nu) \Gamma (k/n)}{\Gamma(1+k/n-\nu)} \Bigl[ F(1-\nu, k/n, 1, x) F(1-\nu, k/n, 1+k/n-\nu, 1- \bar{x}) \\ 
& + F(1-\nu, k/n, 1, \bar{x}) F(1-\nu, k/n, 1+k/n-\nu, 1- x) \Bigr] \\ 
&+ \pi \frac{\sin \pi (k/n-\nu)}{\sin (\pi k/n) \sin \pi \nu}  F(1-\nu, k/n, 1, x)  F(1-\nu, k/n, 1, \bar{x}) \Bigr)^{-1} \Bigr] , \label{ree2}
\end{split}
\end{equation}
where $F(\alpha, \beta, \gamma, z)$ is the Gaussian hypergeometric function,  
\begin{equation}
\begin{split}
F(\alpha, \beta, \gamma, z) = \frac{\Gamma ( \gamma) }{ \Gamma( \beta) \Gamma( \gamma- \beta) } \int_0^1 t^{\beta-1} (1-t)^{\gamma -\beta-1} (1-t z)^{-\alpha} d t ,
\end{split}
\end{equation}
%and we used $\kappa_n^2=(1/\epsilon)^{-2 \frac{k}{n} \left(1-\frac{k}{n} \right)}$, 
%here $\epsilon \equiv a/L$ and $a$ is the UV cutoff length. 
and we used (\ref{I id}) in the second equality and (\ref{I expression}) in the third equality in (\ref{ree2}). 

We study properties of the R\'{e}nyi entropy (\ref{ree2}). 
From (\ref{ree2}), when $\nu \to 0$, $S_A^{(n)}$ diverges as
\begin{equation}
S_A^{(n)} \simeq \ln (1/\nu). 
\end{equation}
This divergence does not depend on the length of the subsystem, 
so it is the contribution of the homogeneous mode. 
This divergence is similar to the infrared divergence of the entanglement entropy in the  massless limit in a free  massive scalar field \cite{CaHu2} and has the similar heuristic explanation.  
The correlation function is given by, 
\begin{equation}
\begin{split}
\langle \phi'(x) \phi'^* (0) \rangle =\frac{L}{2\pi}  e^{-i \nu 2 \pi x /L} \sum_{n=-\infty}^{\infty} \frac{e^{in 2 \pi x /L}}{|n-\nu|} \simeq \frac{L}{2\pi} \frac{1}{\nu} ~~ (\nu \to 0) , \label{two points}
\end{split}
\end{equation}
where we used $w=x+i\tau$ coordinates and $\tau=0$. 
From (\ref{two points}), the typical size of the fluctuations on the homogeneous mode grows as $(1/\nu)^{1/2}$. 
Correspondingly, the  R\'{e}nyi entropy grows as the logarithm of this volume in field space
\cite{Un}, and becomes $S_A^{(n)} \simeq 2 \times  \ln (1/\nu)^{1/2}=\ln (1/\nu)$. 
Note that we doubled the entropy because $\phi'$ is a complex field and has the real part and the imaginary part.

%\begin{figure}
% \includegraphics[width=6.5cm,angle=270,clip]{newthreeregions-1.eps}%
% \caption{Two spheres $A$ and $B$, and the outside region $C$. }
% \label{threeregions}
% \end{figure}

\begin{figure}
\includegraphics[width=6.5cm,clip]{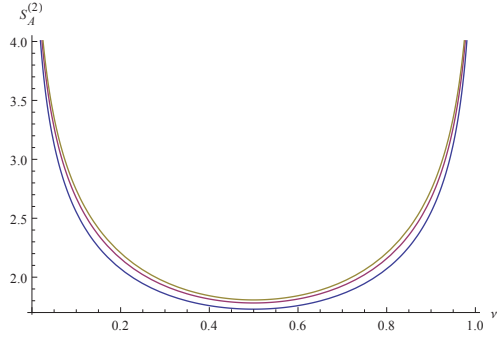}%
\caption{The  R\'{e}nyi entropy $S_A^{(n=2)}$ as a function of $\nu$. From top to bottom:  $l/L=1/6, 1/4, 1/3$.  $S_A^{(n=2)}$ diverges when $\nu \to 0$ and $\nu \to 1$.
$S_A^{(n=2)}$ becomes a minimum value for $\nu=1/2$.
\label{ree2nufig}}
\end{figure}

\begin{figure}
\includegraphics[width=6.5cm,clip]{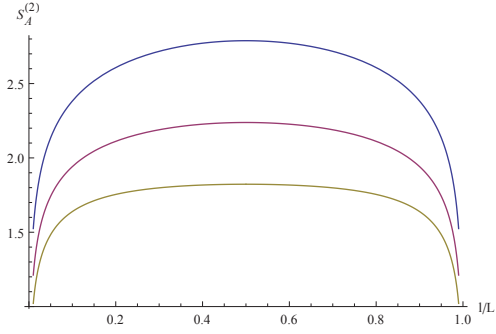}%
\caption{The  R\'{e}nyi entropy $S_A^{(n=2)}$ as a function of $l/L$. From top to bottom:  $\nu=1/10, 1/5, 1/2$.  \label{ree2elfig}}
\end{figure}

We plot the $S_A^{(n=2)}$ as a function of $\nu$ and the length of the subsystem $l$ ( $|x|=2|\sin \frac{\pi l}{L}|$ ) in Fig.\ref{ree2nufig} and Fig.\ref{ree2elfig}.  
In these figures, we have set $\kappa=1$ and $\epsilon=1$. 
The  R\'{e}nyi entropy  $S_A^{(n=2)}$ diverges when $\nu \to 0$ and $\nu \to 1$ as shown in Fig.\ref{ree2nufig}.  
$S_A^{(n=2)}$ becomes a minimum value when $\nu=1/2$.

It is difficult to perform the analytical continuation of the  R\'{e}nyi entropy and to obtain the entanglement entropy because of the complexity of the expression (\ref{ree2}). 
However, we can perform the analytical continuation in the limit $|x| \to 0$ and $\nu \to 0$. 
From (\ref{ree2}), when $|x| \to 0$, we obtain
\begin{equation}
\begin{split}
S_A^{(n)} & \simeq \frac{1}{1-n} \sum_{k=1}^{n-1} \ln  \Bigl[
\left( \frac{|x|}{\epsilon} \right)^{-2 \frac{k}{n} \left(1-\frac{k}{n} \right)} 
 \Bigl( \ln \delta (k/n) + \ln \delta (\nu) - 2 \ln |x|    \Bigr)^{-1} \Bigr]  \\
& = \frac{1}{3} \left(1+\frac{1}{n} \right) \ln \frac{|x|}{\epsilon} +\frac{1}{n-1} \sum_{k=1}^{n-1} \ln 
 \Bigl( \ln \delta (k/n) + \ln \delta (\nu) - 2 \ln |x|    \Bigr)   ,     \label{ree3}
\end{split}
\end{equation}
where
\begin{equation}
\ln \delta (y) =2 \psi (1) -\psi (y) -\psi (1-y),
\end{equation}
here $\psi (y)$ is the digamma function, $\psi (y)=\frac{d}{dy} \ln \Gamma(y)$, 
and we omitted the irrelevant constant $-\ln \kappa^2$. 
From (\ref{ree3}), when $|x| \to 0$ and $\nu \to 0$, we obtain
\begin{equation}
S_A^{(n)} \simeq \frac{1}{3} \left(1+\frac{1}{n} \right) \ln \frac{|x|}{\epsilon} +\ln \left( \frac{1}{\nu} + \ln \frac{1}{|x|^2} \right).
\end{equation}
The first term is the same as the R\'{e}nyi entropy for a free massless complex scalar field and the second term is the correction from the AB phase. 
Thus, when $|x| \to 0$ and $\nu \to 0$, we obtain the entanglement entropy,  
\begin{equation}
S_A^{(n=1)} \simeq \frac{2}{3} \ln \frac{|x|}{\epsilon} +\ln \left( \frac{1}{\nu} + \ln \frac{1}{|x|^2} \right) .  \label{ee1}
\end{equation}

%\section{Free Dirac field}

\section{Conclusion}
We studied the dependence of entanglement entropy with the AB phase in (1+1) dimensional conformal field theory on a one dimensional ring. 
We performed the gauge transformation (\ref{gauge}) and the effect of AB phase is represented by the twisted boundary condition of the scalar field (\ref{bc1}).
We used the conformal map and the boundary condition was expressed by inserting twist operators $\sigma_\nu$ and $\sigma_{1-\nu}$ at $z=0$ and $z=\infty$ in (\ref{vev z}). 
We calculated exactly the R\'{e}nyi entropy in charged free scalar field theory in (\ref{ree2}). 
The  R\'{e}nyi entropy diverges when $\nu \to 0$. 
This divergence comes from the homogeneous mode and is similar to the infrared divergence of the entanglement entropy in a free massive scalar field. 
We gave the heuristic explanation of this divergence. 
We performed the analytical continuation in the limit $|x| \to 0$ and $\nu \to 0$ and obtained the entanglement entropy in (\ref{ee1}).

We considered the ground state in the presence of the AB phase (i.e. the Wilson loop).   
This state is a kind of excited states in CFT without the AB phase. 
Entanglement entropy has been studied to quantify excited states in \cite{NNT, HNTW,  ABS, Sh5}. %No, HNTW, CNT, ABS, Sh5}. 
%By using our method that the effect of AB phase is expressed by inserting twist operators, 
%it is possible to analyze more complicated case, e.g. mutual information of two intervals, time dependent problems and excited states in the presence of the AB phase.  
It is an interesting future problem to apply our method that the effect of AB phase is expressed by inserting twist operators to %more complicated case, e.g. mutual information of two intervals, 
time dependent problems and excited states in the presence of the AB phase.

\begin{acknowledgments}
I am grateful to Daniel Jafferis, Tadashi Takayanagi, and Erik Tonni for useful
comments and discussions. I also would like to thank Masahiro Nozaki, Tokiro Numasawa, and Kento Watanabe for useful discussions. 
I also would like to thank William Witczak-Krempa for letting me know some references about entanglement entropy in quantum field theories with twisted boundary conditions and useful comments on the UV cutoff dependence and the short length and the small AB phase limit of the R\'{e}nyi entanglement entropy.  
This work is supported by Grant-in-Aid for the JSPS Fellowship No.15J02740. 
\end{acknowledgments}

\appendix
\section{The calculation of the integral $I(a, b, c, x)$}

We calculate the integral  $I(a, b, c, x)$ in (\ref{I1}). 
The integral over the complex plane can be evaluated by splitting it into the sum of products of holomorphic and antiholomorphic contour integrals around cuts using a method used in Kawai et al.\cite{KLT},  
\begin{equation}
\begin{split}
& I (a, b, c, x) = \int d^2 z |z|^{2a}  |z-x|^{2b}  |z-1|^{2c}  \\
&=\frac{\sin \pi c \sin \pi a}{\sin \pi (a+b+c)} \Bigl[ \int_0^x d\xi A \int_{\bar{x}}^1 d\eta B +\int_x^1 d\xi A \int_{0}^{\bar{x}} d\eta B \Bigr] \\ 
&+ \frac{\sin \pi (b+ c) \sin \pi a}{\sin \pi (a+b+c)} \int_0^x d\xi A \int_{0}^{\bar{x}} d\eta B + \frac{\sin \pi c \sin \pi ( a+b) }{\sin \pi (a+b+c)}  \int_x^1 d\xi A \int_{\bar{x}}^{1} d\eta B \\
&=\frac{\sin \pi c \sin \pi a}{\sin \pi (a+b+c)} \Bigl[ x^{a+b+1} \frac{\Gamma (a+1) \Gamma(b+1) }{\Gamma(a+b+2)} F(-c, a+1, a+b+2,x) \\
& \times (1-\bar{x})^{1+b+c} \frac{\Gamma (1+c) \Gamma (1+b) }{ \Gamma (2+b+c) } F(-a,1+c, 2+b+c, 1-\bar{x}) + \mathrm{c.c} \Bigr]  \\
& + \frac{\sin \pi (b+ c) \sin \pi a}{\sin \pi (a+b+c)} (x \bar{x} )^{a+b+1}  \left( \frac{\Gamma (a+1) \Gamma (b+1)}{\Gamma (a+b+2)} \right)^2 \\
& \times F(-c, a+1, a+b+2, x) F(-c, a+1,a+b+2, \bar{x} )  \\
& + \frac{\sin \pi c \sin \pi ( a+b) }{\sin \pi (a+b+c)}  ((1-x) (1- \bar{x}) )^{1+b+c} \left( \frac{\Gamma (c+1) \Gamma (b+1)}{\Gamma (b+c+2)} \right)^2 \\
& \times F(-a, c+1, b+c+2, 1-x) F(-a, c+1, b+c+2, 1-\bar{x} ),  \\  \label{I2}
\end{split}
\end{equation}
where
\begin{equation}
\begin{split}
A \equiv |\xi|^a |\xi-x|^b |\xi-1|^c, ~~~ B \equiv  |\eta|^a |\eta-\bar{x}|^b |\eta-1|^c .
\end{split}
\end{equation}
Note that %different expression is written in the Appendix in \cite{Kn}. 
the expression of $I(a, b, c, x)$ in the Appendix in \cite{Kn} is not useful when $a+b=-1$ and we gave the different expression which is useful when $a+b=-1$ in (\ref{I2}).  
We can see that (\ref{I2}) is the same as the result in the Appendix in \cite{Kn} by using the following identity which is obtained by a contour integral around cuts;  
\begin{equation}
\begin{split}
\sin \pi a \int_0^{x} d \xi A+ \sin \pi (a+b) \int_x^1 d \xi A +\sin \pi (a+b+c) \int_1^{\infty} d \xi A =0 .
\end{split}
\end{equation}

From (\ref{I2}), we obtain the necessary integral for the R\'{e}nyi entropy (\ref{ree1}),
\begin{equation}
\begin{split}
 I (a-1, -a, c-1, x) =& \pi \frac{\Gamma (1-c) \Gamma (a)}{\Gamma(1+a-c)} \Bigl[ F(1-c, a, 1, x) F(1-c, a,1+a-c, 1- \bar{x}) \\ 
& + F(1-c, a, 1, \bar{x}) F(1-c, a,1+a-c, 1- x) \Bigr] \\ 
&+ \pi^2 \frac{\sin \pi (a-c)}{\sin \pi a \sin \pi c}  F(1-c, a, 1, x)  F(1-c, a, 1, \bar{x}) , \label{I expression}
\end{split}
\end{equation}
\begin{equation}
\begin{split}
 I (-a, a-1, -c, x) = |1-x|^{2(a-c)} I (a-1, -a, c-1, x) . \label{I id}
\end{split}
\end{equation}

\section{Derivation of the four point function of twist operators by another method}
We calculate the four point function of twist operators by the method in \cite{DFMS, BR}. 
We consider a complex field $X(z, \bar{z})$ and the action for $X(z, \bar{z})$ is given by
\begin{equation}
S[X, \bar{X}] =\frac{1}{4\pi} \int (\partial_z X \partial_{\bar{z}} \bar{X} +\partial_{\bar{z}} X \partial_{z} \bar{X}  ) d^2 z .
\end{equation}
We consider the following four point function
\begin{equation}
Z(z_i, \bar{z_i}) \equiv \langle \mathcal{\sigma}_{k_1/n} (z_1)  \mathcal{\sigma}_{1-k_1/n} (z_2) 
 \mathcal{\sigma}_{k_3/n} (z_3)  \mathcal{\sigma}_{1-k_3/n} (z_4) \rangle  .
\end{equation}

From \cite{DFMS, BR}, we consider the Green function in the presence of four twist operators, 
\begin{equation}
g(z, w, z_i) \equiv \frac{ -\tfrac{1}{2} \langle \partial_z X(z) \partial_w \bar{X}(w)  \mathcal{\sigma}_{k_1/n} (z_1)  \mathcal{\sigma}_{1-k_1/n} (z_2) 
 \mathcal{\sigma}_{k_3/n} (z_3)  \mathcal{\sigma}_{1-k_3/n} (z_4) \rangle}{ \langle \mathcal{\sigma}_{k_1/n} (z_1)  \mathcal{\sigma}_{1-k_1/n} (z_2) 
 \mathcal{\sigma}_{k_3/n} (z_3)  \mathcal{\sigma}_{1-k_3/n} (z_4) \rangle } .
\end{equation}
The Green function obeys the following asymptotic conditions; 
\begin{equation}
\begin{split}
g(z, w, z_i) & \sim (z-w)^{-2} + \textrm{finite} ~~~ \textrm{as}~ z \to w  \\
& \sim \textrm{const} \times (z-z_{1,3})^{-k_{1,3}/n} ~~~ \textrm{as}~ z \to z_{1,3}  \\
& \sim \textrm{const} \times (z-z_{2,4})^{-(1-k_{1,3}/n)} ~~~ \textrm{as}~ z \to z_{2,4}  \\
& \sim \textrm{const} \times (w-z_{1,3})^{-(1-k_{1,3}/n)} ~~~ \textrm{as}~ w \to z_{1,3}  \\
& \sim \textrm{const} \times (w-z_{2,4})^{-k_{1,3}/n} ~~~ \textrm{as}~ w \to z_{2,4} . \\
\end{split}
\end{equation}
Thus, we can write $g(z, w, z_i)$ as
\begin{equation}
\begin{split}
g(z, w, z_i) = \omega_{k} (z) \omega_{n-k} (w) \frac{1}{(z-w)^2} [ A_0(w)+A_1(w) (z-w) +A_2(w) (z-w)^2 ], 
\end{split}
\end{equation}
where 
\begin{equation}
\begin{split}
& \omega_{k} (z) \equiv (z-z_{1})^{-k_{1}/n}   (z-z_{2})^{-(1-k_{1}/n)} (z-z_{3})^{-k_{3}/n} (z-z_{4})^{-(1-k_{3}/n)} \\
& \omega_{n-k} (z) \equiv (z-z_{1})^{-(1-k_{1}/n)}   (z-z_{2})^{-k_{1}/n} (z-z_{3})^{-(1-k_{3}/n)} (z-z_{4})^{-k_{3}/n} \\
\end{split}
\end{equation}
and 
\begin{equation}
\begin{split}
A_0(w) \equiv \prod_{j=1}^4 (w-z_j), ~~~ A_1(w) \equiv A_0(w) \sum_{j=1}^4 \frac{k_j}{n} \frac{1}{w-z_j} , %\left[ \frac{k_1}{n} \frac{1}{w-z_1} +\left(1- \frac{k_1}{n} \right) \frac{1}{w-z_2} \right]
\end{split}
\end{equation}
here we defined $k_{2,4}\equiv n-k_{1,3}$ and $A_2(w)$ will be determined by the  global monodromy condition.   %is a constant in $z$ and $w$.  
The global monodromy condition is 
\begin{equation}
\begin{split}
\Delta_{\mathcal{C}} X = \oint_{\mathcal{C}} dz \partial_z X + \oint_{\mathcal{C}} d\bar{z} \partial_{\bar{z}} X =0  \label{global}
\end{split}
\end{equation}
for all closed loops. 

Before determining $A_2$, we will extract the differential equation of $Z(z_i, \bar{z_i})$. 
Let us consider the limit $w \to z$ 
\begin{equation}
\begin{split}
& \lim_{w \to z} [ g(z, w, z_i) - (z-w)^{-2} ] =  \frac{  \langle  T(z)  \mathcal{\sigma}_{k_1/n} (z_1)  \mathcal{\sigma}_{1-k_1/n} (z_2) 
 \mathcal{\sigma}_{k_3/n} (z_3)  \mathcal{\sigma}_{1-k_3/n} (z_4) \rangle}{ \langle \mathcal{\sigma}_{k_1/n} (z_1)  \mathcal{\sigma}_{1-k_1/n} (z_2) 
 \mathcal{\sigma}_{k_3/n} (z_3)  \mathcal{\sigma}_{1-k_3/n} (z_4) \rangle }  \\
&= A_2 (z) \sum_j \frac{1}{z-z_j} \prod_{l \neq j} (z_j-z_l)^{-1} + \sum_{j=1}^4 h_j (z-z_j)^{-2} - \sum_{j} \frac{1}{z-z_j} \sum_{l \neq j} \frac{k_j}{n} \frac{k_l}{n} \frac{1}{z_j-z_l}  , \label{limit1}
\end{split}
\end{equation}
where $T(z)$ is the stress tensor and $h_j \equiv \frac{1}{2} (k_j/n) \left(1-k_j/n \right)$ is the conformal weight of the twist operator $\sigma_{k_j/n}$. 
We apply the operator product
\begin{equation}
\begin{split}
T (z) \mathcal{\sigma}_{k_2} (z_2) \sim \frac{h_2 \mathcal{\sigma}_{k_2} (z_2) }{(z-z_2)^2} + \frac{\partial_{z_2} \mathcal{\sigma}_{k_2} (z_2) }{z-z_2} + \dots
\end{split}
\end{equation}
to (\ref{limit1}) and obtain the differential equation
\begin{equation}
\begin{split}
& \partial_{z_2} \ln Z(z_i, \bar{z_i}) = \frac{A_2 (z_2)}{ (z_2-z_1)(z_2-z_3)(z_2-z_4)} -\frac{k_2}{n} \left( \frac{k_1}{n} \frac{1}{z_2-z_1} +\frac{k_3}{n} \frac{1}{z_2-z_3}+\frac{k_4}{n} \frac{1}{z_2-z_4} \right) . \label{differential1}
\end{split}
\end{equation}
It is useful to use the following conformal map
\begin{equation}
\begin{split}
z \to \frac{(z_1-z)(z_3-z_4)}{(z_1-z_3)(z-z_4)},
\end{split}
\end{equation}
which sends $z_1,~ z_2,~ z_3$, and $ z_4$ into $0,~x,~1$ and $\infty$ respectively, 
where $x$ is the cross ratio ($z_{ij}\equiv z_i-z_j$)
\begin{equation}
\begin{split}
x \equiv \frac{z_{12}z_{34}}{z_{13}z_{24}}, ~~~ x(1-x) = \frac{z_{12}z_{34}z_{14}z_{23}}{z_{13}^2 z_{24}^2} .
\end{split}
\end{equation}
Thus (\ref{differential1}) becomes
\begin{equation}
\begin{split}
& \partial_{x} \ln Z(x, \bar{x}) = - \frac{\tilde{A_2} }{ x(1-x)} -\frac{k_2}{n} \left( \frac{k_1}{n} \frac{1}{x} -\frac{k_3}{n} \frac{1}{1-x} \right) . \label{differential2}
\end{split}
\end{equation}
where 
\begin{equation}
\begin{split}
&  Z(x, \bar{x}) = \lim_{z_{\infty} \to \infty}  \langle \mathcal{\sigma}_{k_1/n} (0)  \mathcal{\sigma}_{1-k_1/n} (x) 
 \mathcal{\sigma}_{k_3/n} (1)  \mathcal{\sigma}_{1-k_3/n} (z_\infty) \rangle   \\
& \tilde{A_2} = \lim_{z_{\infty} \to \infty}  -z_{\infty}^{-1} A_{2} (z_1=0, z_2=x, z_3=1, z_4=z_\infty; w=x) .
\end{split}
\end{equation}

\begin{figure}
\includegraphics[width=10cm,clip]{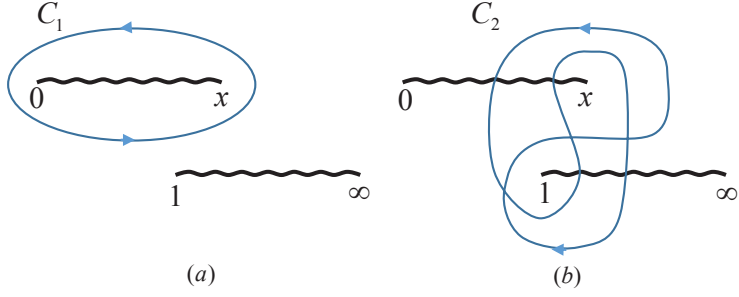}%
\caption{The two closed loops $\mathcal{C}_1$ and $\mathcal{C}_2$ we consider as a basis of the loops in the complex plane.  \label{ABloop}}
\end{figure}

In order to determine $A_2$ by the  global monodromy condition, we introduce the auxiliary correlation function 
\begin{equation}
\begin{split}
h(z, w, z_i) & \equiv \frac{ -\tfrac{1}{2} \langle \partial_{\bar{z}} X(\bar{z}) \partial_w \bar{X}(w)  \mathcal{\sigma}_{k_1/n} (z_1)  \mathcal{\sigma}_{1-k_1/n} (z_2) 
 \mathcal{\sigma}_{k_3/n} (z_3)  \mathcal{\sigma}_{1-k_3/n} (z_4) \rangle}{ \langle \mathcal{\sigma}_{k_1/n} (z_1)  \mathcal{\sigma}_{1-k_1/n} (z_2) 
 \mathcal{\sigma}_{k_3/n} (z_3)  \mathcal{\sigma}_{1-k_3/n} (z_4) \rangle } \\
& = B(z_i, \bar{z}_i) \bar{\omega}_{n-k}(\bar{z}) \omega_{n-k} (w),
\end{split}
\end{equation}
where we determined $h$ in the same way as $g$ was determined. 
From the global monodromy condition (\ref{global}), we obtain
\begin{equation}
\begin{split}
 \oint_{\mathcal{C}_i} dz g(z, w, z_i) + \oint_{\mathcal{C}_i} d\bar{z} h(z, w, z_i) =0 , ~~~ i=1,2 \label{global2}
\end{split}
\end{equation}
where we chose the two loops $\mathcal{C}_1$ and $\mathcal{C}_2$ shown in Fig.\ref{ABloop}  as a basis of the loops.

We divide (\ref{global2}) by $\omega_{n-k} (w)$ and set $w=z_2$, $z_1=0$, $z_2=x$, $z_3=1$ and $z_4 \to \infty$ and obtain
\begin{equation}
\begin{split}
\left( -x(1-x) \frac{d}{dx} +\tilde{A_2} \right)  \oint_{\mathcal{C}_i} dz \omega_{k}' (z) + \tilde{B} \oint_{\mathcal{C}_i} d\bar{z} \bar{\omega}_{n-k}'(\bar{z}) =0 , ~~~ i=1,2 \label{global3}
\end{split}
\end{equation}
where $\tilde{B}\equiv \lim_{z_{\infty} \to \infty} z_\infty^{-2} B (z_1=0, z_2=x, z_3=1, z_4=z_\infty) $ and 
\begin{equation}
\begin{split}
& \omega_{k}' (z) \equiv z^{-k_{1}/n}   (z-x)^{-(1-k_{1}/n)} (z-1)^{-k_{3}/n}  \\
&\bar{ \omega}_{n-k}' (\bar{z}) \equiv \bar{z}^{-(1-k_{1}/n)}   (\bar{z}-\bar{x})^{-k_{1}/n} (\bar{z}-1)^{-(1-k_{3}/n)} .  \\
\end{split}
\end{equation} 
We calculate all integrals in (\ref{global3}) and obtain
\begin{equation}
\begin{split}
 \oint_{\mathcal{C}_1} dz \omega_{k}' (z) =(-1+ \alpha^{-k_1}) \alpha^{-\frac{1}{2}(k_2+k_3)} \Gamma \left(1-\frac{k_1}{n} \right) \Gamma \left( \frac{k_1}{n} \right) F \left( \frac{k_3}{n}, 1-\frac{k_1}{n}, 1, x \right),   
\end{split}
\end{equation}
\begin{equation}
\begin{split}
& \oint_{\mathcal{C}_1} d\bar{z} \bar{\omega}_{n-k}'(\bar{z}) =(-1+ \alpha^{-k_1}) \alpha^{-\frac{1}{2}(k_2+k_3)} \Gamma \left(1-\frac{k_1}{n} \right) \Gamma \left( \frac{k_1}{n} \right) F \left(1-\frac{k_3}{n}, \frac{k_1}{n}, 1, \bar{x} \right),   \\
\end{split}
\end{equation}
\begin{equation}
\begin{split}
 \oint_{\mathcal{C}_2} dz \omega_{k}' (z) =& (-1+\alpha^{-k_2} -\alpha^{-(k_2+k_3)} +\alpha^{-k_3})  \alpha^{-\frac{1}{2}k_3}  (1-x)^{\frac{(k_1-k_3)}{n} } \\ 
& \times F \left( \frac{k_1}{n}, 1-\frac{k_3}{n}, 1-\frac{k_3}{n} +\frac{k_1}{n}, 1-x \right) %\\
%& \times 
\frac{\Gamma \left(1-\frac{k_3}{n} \right) \Gamma \left( \frac{k_1}{n} \right) }{\Gamma \left( 1-\frac{k_3}{n} +\frac{k_1}{n} \right)},    \\
\end{split}
\end{equation}
\begin{equation}
\begin{split}
 \oint_{\mathcal{C}_2} d\bar{z} \bar{\omega}_{n-k}'(\bar{z}) =& -  (-1+\alpha^{-k_2} -\alpha^{-(k_2+k_3)} +\alpha^{-k_3})  \alpha^{-\frac{1}{2}k_3} (1-\bar{x})^{\frac{(k_3-k_1)}{n} } \\ 
& \times F \left(1-\frac{k_1}{n}, \frac{k_3}{n}, 1+\frac{k_3}{n} -\frac{k_1}{n}, 1-\bar{x} \right) %\\ 
%& \times 
\frac{\Gamma \left(1-\frac{k_1}{n} \right) \Gamma \left( \frac{k_3}{n} \right) }{\Gamma \left( 1+\frac{k_3}{n} -\frac{k_1}{n} \right)},    \\
\end{split}
\end{equation}
where $\alpha \equiv e^{i 2\pi /n}$. 
Solving equations (\ref{global3}) for $\tilde{A_2}$, we obtain 
\begin{equation}
\begin{split}
\frac{\tilde{A_2}}{x(1-x)} = \partial_x \ln \left[ \oint_{\mathcal{C}_1} dz \omega_{k}' (z)  \oint_{\mathcal{C}_2} d\bar{z} \bar{\omega}_{n-k}'(\bar{z}) - \oint_{\mathcal{C}_2} dz \omega_{k}' (z) \oint_{\mathcal{C}_1} d\bar{z} \bar{\omega}_{n-k}'(\bar{z}) \right] . \label{det}
\end{split}
\end{equation}
%From (\ref{differential2}) and (\ref{det})
We substitute (\ref{det}) for (\ref{differential2}) and obtain
\begin{equation}
\begin{split}
Z(x, \bar{x}) =C(\bar{x}) x^{-k_1 k_2/n^2} (1-x)^{-k_2 k_3/n^2} (W(k_1/n, k_3/n, x, \bar{x}))^{-1} , \label{Zx}
\end{split}
\end{equation}
where $C(\bar{x})$ is an arbitrary function of $\bar{x}$ and 
\begin{equation}
\begin{split}
&W(k_1/n, k_3/n, x, \bar{x}) \\ 
&= \frac{\Gamma \left(1-\frac{k_1}{n} \right) \Gamma \left(\frac{k_3}{n} \right) }{\Gamma \left( 1+\frac{k_3}{n} -\frac{k_1}{n} \right)} (1-\bar{x})^{\frac{(k_3-k_1)}{n}} F \left(1-\frac{k_1}{n}, \frac{k_3}{n}, 1+\frac{k_3}{n} -\frac{k_1}{n}, 1-\bar{x} \right) F\left( \frac{k_3}{n}, 1-\frac{k_1}{n}, 1, x\right) \\
&+  \frac{\Gamma \left(1-\frac{k_3}{n} \right) \Gamma \left(\frac{k_1}{n} \right) }{\Gamma \left( 1+\frac{k_1}{n} -\frac{k_3}{n} \right)} (1-x)^{\frac{(k_1-k_3)}{n}} F \left(\frac{k_1}{n}, 1-\frac{k_3}{n}, 1+\frac{k_1}{n} -\frac{k_3}{n}, 1-x \right) F\left( 1-\frac{k_3}{n}, \frac{k_1}{n}, 1, \bar{x} \right)  .
\end{split}
\end{equation}
In order to fix the $\bar{x}$-dependence of $Z(x, \bar{x})$, we consider $\partial_{\bar{x}} \ln Z(x, \bar{x})$ in the same way as $\partial_{x} \ln Z(x, \bar{x})$. 
The differential equation for  $\partial_{\bar{x}} \ln Z(x, \bar{x})$ is obtained by replacing $x \to \bar{x}$, $\bar{x} \to x$ and  $k_i \to n- k_i$ in that for $\partial_{x} \ln Z(x, \bar{x})$ and we obtain
\begin{equation}
\begin{split}
Z(x, \bar{x}) =D(x) \bar{x}^{-k_1 k_2/n^2} (1-\bar{x})^{-(n-k_2) (n-k_3)/n^2} (W(1-k_1/n, 1-k_3/n, \bar{x}, x))^{-1} , \label{Zxbar}
\end{split}
\end{equation}
where $D(x)$ is an arbitrary function of $x$. 
Finally, from (\ref{Zx}) and (\ref{Zxbar}), we obtain
\begin{equation}
\begin{split}
Z(x, \bar{x}) =f \cdot |x|^{-2k_1 (n-k_1)/n^2} |1-x|^{2 k_1 k_3/n^2-(k_1+k_3)/n} ( \tilde{W}(k_1/n, k_3/n, x, \bar{x}))^{-1} , \label{Z final}
\end{split}
\end{equation}
where $f$ is an integration constant and  
\begin{equation}
\begin{split}
& \tilde{W}(k_1/n, k_3/n, x, \bar{x}) \\ 
&= \frac{\Gamma \left(1-\frac{k_1}{n} \right) \Gamma \left(\frac{k_3}{n} \right) }{\Gamma \left( 1+\frac{k_3}{n} -\frac{k_1}{n} \right)} |1-x|^{\frac{(k_3-k_1)}{n}} F \left(1-\frac{k_1}{n}, \frac{k_3}{n}, 1+\frac{k_3}{n} -\frac{k_1}{n}, 1-\bar{x} \right) F\left( \frac{k_3}{n}, 1-\frac{k_1}{n}, 1, x\right) \\
&+  \frac{\Gamma \left(1-\frac{k_3}{n} \right) \Gamma \left(\frac{k_1}{n} \right) }{\Gamma \left( 1+\frac{k_1}{n} -\frac{k_3}{n} \right)} |1-x|^{\frac{(k_1-k_3)}{n}} F \left(1-\frac{k_3}{n},  \frac{k_1}{n}, 1+\frac{k_1}{n} -\frac{k_3}{n}, 1-x \right) F\left( 1-\frac{k_3}{n}, \frac{k_1}{n}, 1, \bar{x} \right)  .
\end{split}
\end{equation}
We can show that the integration constant $f$ is independent of $k_1$ and $k_3$ by taking the limit $|x| \to 0$ and considering the OPE of $\sigma_{k_1/n} (0) \sigma_{1-k_1/n} (x)$. 
After some calculation, we obtain the following identity; 
\begin{equation}
\begin{split}
& (\tilde{W}(a, c, x, \bar{x}))^2 = \frac{1}{\pi^2} I(a-1, -a, c-1, x) I(-a, a-1, -c, x). 
\end{split}
\end{equation}
Thus (\ref{Z final}) is equal to (\ref{four point 1}) for $k_2=n-k_1$ when we set $f=\kappa^2/\pi$.

%\begin{acknowledgments}
%\end{acknowledgments}


\begin{thebibliography}{99}











\bibitem{RT}
  S.~Ryu and T.~Takayanagi,
  ``Holographic derivation of entanglement entropy from AdS/CFT,''
  Phys.\ Rev.\ Lett.\  {\bf 96} (2006) 181602;
  %%CITATION = PRLTA,96,181602;%%
 ``Aspects of holographic entanglement entropy,''
  JHEP {\bf 0608} (2006) 045.
  %%CITATION = JHEPA,0608,045;%%
  

\bibitem{Fa}
T. Faulkner, A. Lewkowycz and J. Maldacena, ``Quantum corrections to holographic
entanglement entropy," JHEP 1311, 074 (2013) [arXiv:1307.2892].



\bibitem{Sw}
B. Swingle, "Entanglement Renormalization and Holography," Phys. Rev. D 86, 065007
(2012), arXiv:0905.1317 [cond-mat.str-el].


%\bibitem{GMT}
%K. Goto, M. Miyaji, T. Takayanagi, "Causal Evolutions of Bulk Local Excitations from CFT, " JHEP 1609 (2016) 130
%arXiv:1605.02835 [hep-th]. 

\bibitem{NRT}
M. Nozaki, S. Ryu and T. Takayanagi, "Holographic Geometry of Entanglement Renormalization in Quantum Field Theories," JHEP 1210 (2012) 193, arXiv:1208.3469
[hep-th].

\bibitem{MT}
M. Miyaji and T. Takayanagi, "Surface/State Correspondence as a Generalized Holography," PTEP 2015 (2015) no.7, 073B03, 
arXiv:1503.03542[hep-th]. 

\bibitem{RaT}
M. Rangamani, T. Takayanagi, "Holographic Entanglement Entropy, " arXiv:1609.01287 [hep-th].

%\bibitem{MTW}
%M. Miyaji, T. Takayanagi, K. Watanabe, "From Path Integrals to Tensor Networks for AdS/CFT, " arXiv:1609.04645 [hep-th].

\bibitem{Sh3}
N. Shiba, T. Takayanagi, ``Volume Law for the Entanglement Entropy in Non-local QFTs,'' JHEP 1402 (2014) 033,
arXiv:1311.1643 [hep-th].

\bibitem{Sh4}
A. Mollabashi, N. Shiba, T. Takayanagi, ``Entanglement between Two Interacting CFTs and Generalized Holographic Entanglement Entropy,'' JHEP 1404 (2014) 185,
arXiv:1403.1393 [hep-th]. 

\bibitem{Sh7}
M. Miyaji, T. Numasawa, N. Shiba, H. Takayanagi, K. Watanabe, ``Continuous Multiscale Entanglement Renormalization Ansatz as Holographic Surface-State Correspondence,'' Phys.Rev.Lett. 115 (2015) no.17, 171602,
arXiv:1506.01353 [hep-th].

\bibitem{Sh8}
M. Miyaji, T. Numasawa, N. Shiba, H. Takayanagi, K. Watanabe, ``Distance between Quantum States and Gauge-Gravity Duality,'' Phys.Rev.Lett. 115 (2015) no.26, 261602,
arXiv:1507.07555 [hep-th].

\bibitem{Sh9}
T. Miyagawa, N. Shiba, H. Takayanagi, ``Double-Trace Deformations and Entanglement Entropy in AdS,'' Fortsch.Phys. 64 (2016) 92-105,
 arXiv:1511.07194 [hep-th].

\bibitem{Sh10}
T. Numasawa, N. Shiba, H. Takayanagi, K. Watanabe, ``EPR Pairs, Local Projections and Quantum Teleportation in Holography,'' JHEP 1608 (2016) 077,
arXiv:1604.01772 [hep-th].





\bibitem{LW}
M. Levin and X.-G. Wen, “Detecting Topological Order in a Ground State Wave Function,” Phys. Rev. Lett. 96, 110405 (2006), arXiv:cond-mat/0510613.



\bibitem{KP}
A. Kitaev and J. Preskill, “Topological entanglement entropy,” Phys. Rev. Lett. 96, 110404 (2006), 
arXiv:hep-th/0510092


\bibitem{CC}
P. Calabrese and J. Cardy, “Entanglement entropy and quantum field theory,” J. Stat. Mech. 0406, P002 (2004), arXiv:hep-th/0405152


\bibitem{GST}
S. Ghosh, R. M. Soni, S. P. Trivedi, "On The Entanglement Entropy For Gauge Theories, " 
JHEP 1509 (2015) 069 ,  arXiv:1501.02593 [hep-th]. 

\bibitem{Sh6}
S. Aoki, T. Iritani, M. Nozaki, T. Numasawa, N. Shiba, H. Tasaki, `` On the definition of entanglement entropy in lattice gauge theories,'' JHEP 1506 (2015) 187,
arXiv:1502.04267 [hep-th].




\bibitem{Bombelli:1986rw}
  L.~Bombelli, R.~K.~Koul, J.~Lee and R.~D.~Sorkin,
  ``A Quantum Source of Entropy for Black Holes,''
  Phys.\ Rev.\ D {\bf 34} (1986) 373.
  %% CITATION = PHRVA,D34,373;%%

\bibitem{Sr}
   M.~Srednicki,
   ``Entropy and area,''
   Phys.\ Rev.\ Lett.\  {\bf 71} (1993) 666  [hep-th/9303048].
   %% CITATION = HEP-TH/9303048;%%

\bibitem{SU}
L. Susskind and J. Uglum, "Black hole entropy in canonical quantum gravity and superstring theory," Phys. Rev. D 50, 2700 (1994), arXiv:hep-th/9401070.

\bibitem{Ka}
D. N. Kabat, "Black hole entropy and entropy of entanglement," Nucl. Phys. B 453, 281
(1995), arXiv:hep-th/9503016.

\bibitem{Sh1}
N. Shiba, ``Entanglement Entropy of Two Black Holes and Entanglement Entropic
Force,'' Phys.Rev. D83 (2011) 065002, arXiv:1011.3760 [hep-th].

\bibitem{Sh2}
N. Shiba, ``Entanglement Entropy of Two Spheres,'' JHEP 1207 (2012) 100,
arXiv:1201.4865 [hep-th].





\bibitem{ABC}
R. E. Arias, D. D. Blanco, H. Casini, "Entanglement entropy as a witness of the Aharonov-Bohm effect in QFT," J.Phys. A48 (2015) no.14, 145401, 
arXiv:1409.3269 [hep-th]


\bibitem{CCDSLMV}
 L. Chojnacki, C. Q. Cook, D. Dalidovich, L. E. H. Sierens, É. Lantagne-Hurtubise, R. G. Melko, T. J. Vlaar, "Shape dependence of two-cylinder Renyi entropies for free bosons on a lattice," Phys. Rev. B 94, 165136 (2016), 
 arXiv:1607.05311 [cond-mat.str-el]

\bibitem{MFS}
 M. A. Metlitski, C. A. Fuertes, S. Sachdev
"Entanglement Entropy in the O(N) model," 
Phys. Rev. B 80, 115122 (2009), 
 arXiv:0904.4477 [cond-mat.stat-mech] 

\bibitem{WWS}
S. Whitsitt, W. Witczak-Krempa, S. Sachdev
"Entanglement entropy of the large N  Wilson-Fisher conformal field theory,"
Phys. Rev. B 95, 045148 (2017)
arXiv:1610.06568 [cond-mat.str-el]

\bibitem{CWFF}
X. Chen, W. Witczak-Krempa, T. Faulkner, E. Fradkin
"Two-cylinder entanglement entropy under a twist,"
J.Stat.Mech. 1704 (2017) no.4, 043104 
arXiv:1611.01847 [cond-mat.str-el]






\bibitem{CCT}
P. Calabrese, J. Cardy, E. Tonni, "Entanglement entropy of two disjoint intervals in conformal field theory, "  J.Stat.Mech. 0911 (2009) P11001, arXiv:0905.2069 [hep-th]. 

\bibitem{Kn}
V. G. Knizhnik, "Analytic fields on Riemann surfaces. II," Communn. Math. Phys. 112, 567 (1987).

\bibitem{DFMS}
L. J. Dixon, D. Friedan, E. J. Martinec and S. H. Shenker, "The Conformal Field Theory Of
Orbifolds," Nucl. Phys. B 282 (1987) 13.

\bibitem{BR}
M. Bershadsky and A. Radul, "Conformal field theories with additional $Z_N$ symmetry," Int. J. Mod. Phys. A 2, 165 (1987).

\bibitem{ADGN}
J. J. Atick, L. J. Dixon, P. A. Griffin and D. Nemeschansky, "Multiloops twist field correlation
functions for Z(N) orbifolds," Nucl. Phys. B 298 (1988) 1.




\bibitem{CaHu2}
H. Casini and M. Huerta, ``Entanglement entropy in free quantum field theory,''
J.Phys. A42 (2009) 504007, arXiv:0905.2562 [hep-th].

\bibitem{Un}
W. G. Unruh, "Comment on `Proof of the quantum bound on specific entropy for free fields' ," Phys. Rev. D 42, 3596 (1990).












\bibitem{NNT}
M. Nozaki, T. Numasawa and T. Takayanagi, "Quantum Entanglement of Local Operators in Conformal Field Theories," Phys. Rev. Lett. 112, 111602 (2014)
[arXiv:1401.0539 [hep-th]].



%\bibitem{No}
%M. Nozaki, "Notes on Quantum Entanglement of Local Operators," JHEP 1410 (2014) 147,  arXiv:1405.5875
%[hep-th].


\bibitem{HNTW}
S. He, T. Numasawa, T. Takayanagi and K. Watanabe, "Quantum Dimension as Entanglement
Entropy in 2D CFTs,"  Phys.Rev. D90 (2014) no.4, 041701,  arXiv:1403.0702 [hep-th].

%\bibitem{CNT}
%P. Caputa, M. Nozaki and T. Takayanagi, 
%"Entanglement of Local Operators in large N CFTs," PTEP 2014 (2014) 093B06,
%arXiv:1405.5946 [hep-th]

\bibitem{ABS}
F. C. Alcaraz, M. I. Berganza, G. Sierra, Phys. Rev. Lett.
106 (2011) 201601 [arXiv:1101.2881 [cond-mat]].




\bibitem{Sh5}
N. Shiba, ``Entanglement Entropy of Disjoint Regions in Excited States : An Operator Method,'' JHEP 1412 (2014) 152,
arXiv:1408.0637 [hep-th].



\bibitem{KLT}
H. Kawai, D.C. Lewellen, S.-H. H. Tye, "A Relation Between Tree Amplitudes of Closed and Open Strings," Nucl.Phys. B269 (1986) 1-23.  




\end{thebibliography}

\end{document}